\documentclass[12pt,english]{article}
\usepackage[T1]{fontenc}
\usepackage[latin9]{inputenc}
\usepackage{prettyref}
\usepackage{amsmath}
\usepackage{amssymb}
\usepackage{graphicx}
\usepackage{esint}

\makeatletter

\usepackage{amsthm}\textwidth=18cm\oddsidemargin=-15mm\topmargin=-1cm

\newcommand{\be}{\begin{equation}}\newcommand{\ee}{\end{equation}}\newcommand{\ba}{\begin{array}}\newcommand{\ea}{\end{array}}\newcommand{\bea}{\begin{eqnarray}}\newcommand{\eea}{\end{eqnarray}}

\newcommand{\ket}[1]{|#1\rangle}

\numberwithin{lemma}{section}\numberwithin{corol}{section}\numberwithin{prop}{section}\numberwithin{dfn}{section}\numberwithin{equation}{section}

\usepackage{babel}

\usepackage{babel}

\makeatother

\usepackage{babel}
\begin{document}

\title{Multi-qubit parity measurement in circuit quantum electrodynamics}

\author{David P. DiVincenzo and Firat Solgun%
\thanks{RWTH Aachen and Forschungszentrum Juelich, Germany. \texttt{d.divincenzo@fz-juelich.de/firat.solgun@gmail.com} %
}}
\maketitle
\begin{abstract}
We present a concept for performing direct parity measurements on
three or more qubits in microwave structures with superconducting resonators
coupled to Josephson-junction qubits. We write the quantum-eraser
conditions that must be fulfilled for the parity measurements as requirements
for the scattering phase shift of our microwave structure. We show
that these conditions can be fulfilled with present-day devices. We
present one particular scheme, implemented with two-dimensional cavity
techniques, in which each qubit should be coupled equally to two different
microwave cavities. The magnitudes of the couplings that are needed
are in the range that has been achieved in current experiments. A
quantum calculation indicates that the measurement is optimal if the
scattering signal can be measured with near single photon sensitivity.
A comparison with an extension of a related proposal from cavity optics
is presented. We present a second scheme, for which a scalable implementation
of the four-qubit parities of the surface quantum error correction
code can be envisioned. It uses three-dimensional cavity structures,
using cavity symmetries to achieve the necessary multiple resonant
modes within a single resonant structure. 
\end{abstract}
\newpage{}

\section{Introduction}

The essence of error correcting either quantum or classical information
is parity checking. In all practical quantum error correction codes\cite{Mermin},
the error-free state is signalled by parities of a selection of subsets
of qubits all being ``even''; conversely, the occurrence of ``odd''
parities indicates a non-trivial {\em error syndrome}, with which
the particular form of errors can be diagnosed. Calculations show
that remarkably simple codes are very effective as the substrate of
fault tolerant quantum computation; the subsets subjected to parity
checks are geometrically local on a two-dimensional lattice. In addition,
the weight of the parity checks is low; for the canonical code of
this class, the Kitaev toric code\cite{Dennis}, the weights of all
checks are four.

It has been standard to assume that a qubit parity should be obtained
by computation, in particular by a series of two-qubit quantum logic
gates. Thus for the weight-4 case, the circuit involves four controlled-NOT
gates from each of the four qubits in succession to a fifth, ancilla
qubit. The ancilla then holds the value of the parity (or is, perhaps,
in an entangled superposition of the two different parity states),
so that a measurement of the ancilla reveals (or fixes) this parity.
In this paper, we show that, by use of standard microwave scattering
techniques, the parity of a small subset of qubits may be measured
{\em directly}, without the need of an intermediate calculation
requiring a logic circuit. We hope that this will simplify the process
of error correction, and improve thereby its error robustness.

We first discuss some features of the parity determination that are
of a particularly quantum-mechanical character. First, it cannot be
trivially assumed that after a quantum measurement records some value,
the state of quantum object necessarily still has that value. For
example, if a polarization-sensitive photodetector ``clicks\textquotedbl{}
to indicate the polarization of the photon, the photon possessing
that polarization will have vanished. In quantum mechanics there is
a name for measurements for which the quantum object remains in the
state that is recorded: these are called {\em quantum nondemolition}
(QND) measurements. Fortunately, there are many implementations of
QND-type measurements, and the types of scattering measurements proposed
here will have the QND character.

Second, there is a choice of basis involved in defining parity. Thus,
while in the classical basis(called the ``Z basis''\cite{Mermin})
we would call 0000, 0011, 0110, etc., the even parity states, it is
possible to take the basis of the qubit to be, for example $\ket{+}=(\ket{0}+\ket{1})/\sqrt{2}$
and $\ket{-}=(\ket{0}-\ket{1})/\sqrt{2}$. In this ``X basis'',
the even parity states are $++++$, $++--$, $+--+$, etc. In fact,
both the X- and Z-type parity checks are needed in quantum error correction.
We will introduce a parity measurement for just one basis, understanding
that it is possible apply one-qubit gates to rotate the qubits so
that the parity detection is either of X- or of Z-type. In the superconducting
qubit systems that we discuss, these one-qubit rotations can be performed
very accurately and quickly.

Third, it is crucial that the parity measurement reveal {\em only}
the parity, and nothing more. For example, the states 0000 and 0011
should both give parity ``even'', and should be seen as identical
in the measurement process. This concept has no meaning in the classical
setting, where 0000 and 0011 represent objectively different states.
Quantum mechanics permits a state like $(\ket{0000}+\ket{0011})/\sqrt{2}$,
which does not have a specific, definite bit state, but which nevertheless
has a definite parity. In fact, we employ here a strengthening of
the idea of QND, which traditionally requires only that a quantum
state remain in a certain subspace after measurement. For the measurements
that we need, the state is to remain exactly unchanged, and two states
with the same parity should reveal nothing of the differences between
them.

In quantum physics, this final concept has been given a name, the
{\em quantum eraser}\cite{eraser}. This name refers to the fact
that typically in the course of the measurement process, information
is temporarily imprinted on the measurement probe, which is however
erased by the end of the measurement process. In interferometry this
information is the {\em welcher weg} (``which path'') information
which temporarily exists while a photon is moving through the interferometer.
This information is erased by the passage of the photon through the
final beam splitter of the interferometer. In the implementation we
develop below, the measurement probe will temporarily have ``which
bitstring'' information, which, by the time the scattering
process is complete, will be all erased, except for a single bit of
parity information. We will show explicitly two different microwave
protocols which will permit this {\em quantum eraser condition}
to be satisfied; we will in fact precisely quantify the degree to
which this condition is fulfilled.

Discussions of parity measurements are not new to quantum computing
theory; it was understood almost from the beginning that two-qubit
parity measurement permitted the implementation of standard two-qubit
logic gates\cite{firstparity}. A wide variety of two-qubit implementations
have been proposed, in electron optics\cite{BDEK}, for spin qubits
in quantum dots\cite{lossengel}, and for charge qubits\cite{chargeentangler}.
The previous discussions of two-qubit parity measurement for atoms
in optical cavities\cite{mabuchiway}, and for superconducting qubits\cite{coherent2,Tornberg},
will be particularly relevant for the present work. Other theoretical
work has simulated the details of the parity measurement process \cite{korotkovQND,jmeastheory},
and more generally of ``joint measurement'', in which other two-qubit
operators other than the parity are detected. Such joint measurements
have been achieved in the area of circuit QED\cite{jointYale,jointETHZ,2bitalg,jointinit},
the implementation that we discuss here.

Especially for the application to fault-tolerant error correction\cite{devitt,parityimportant},
it is very important to go beyond two-qubit parity. While quantum
error correction codes are known for which two-qubit parity measurements
do suffice\cite{subsys2}, they are found to have much worse threshold-rate
behavior than the Kitaev code\cite{Dennis}, which requires 4-qubit
checks\cite{RHhigh2}. The Bacon-Shor codes\cite{Bacon} also permit
error correction with only two-bit parity checks, and have rather
good error correction performance\cite{cross}, but cannot achieve
fault-tolerant operation by only local operations\cite{cross2}. Many
architectural details of a quantum computer employing the surface
code have been worked out\cite{DDV}, and the outlook seems quite
favorable\cite{RHhigh2,Fowler,Horsman}. Much less is known about
the efficacy of 3-qubit parity measurements for error correction,
although there is a very promising, recent preliminary result\cite{subsys3}.

Extending existing two-qubit parity measurements to more than two
is not trivial. Classically combining many two-bit parity results
to obtain a multi-qubit parity is not permitted, as the overall quantum
eraser condition is not satisfied in this case.  A real modification
of the measurement protocol or of the coupling structure is necessary. For some other existing schemes\cite{coherent2,Tornberg} we see
no reasonable extension to more than two qubits; in the case of Kerchhoff
{\em et al.}\cite{mabuchiway} such an extension is possible, as
we note below. There is another related proposal for a multiqubit parity measurement in optical systems\cite{Yama} based on the acquisition of successive small phase shifts by a probe beam; this scheme seems to require some fine tuning, and has no clear extension beyond three qubits.  A very recent proposal shows another technique related
to the one we present here for extracting multi-qubit parities\cite{newnigg}.

We present here a new solution, requiring a specifically designed
cavity (or multi-cavity) structure with multiple, closely spaced resonant
modes.  We first provide a detailed proposal for performing the 3-qubit parity measurement,
with accurate satisfaction of the quantum eraser condition, in circuit
quantum electrodynamics using superconducting qubits. In the current
work the preferred type of superconducting qubit is the ``transmon\textquotedbl{}
type\cite{transmon}, although other types would also be possible\cite{CSFQ}.
We will always require that the couplings between qubits and microwave
resonators be in the well-studied ``dispersive regime''\cite{blaischi,blais},
in which the qubit transition frequency and the microwave resonant
frequency are well separated. In this regime, the qubit gives a state-dependent
shift to the resonant frequency of a microwave cavity -- one shift
for qubit state $\ket{0}$, and a different shift for qubit state
$\ket{1}$ (note that state 0 and state 1 are at different energies).
In this regime, the requirements of the multi-qubit parity measurement
reduce to those of a classical microwave design problem.

Our first proposal requires a particular hardware arrangement, involving
a capability that has been developed only in the most recent experimental
literature\cite{johnson,mariantoni}. In particular, we require the
qubits involved in a 3-qubit parity measurement each to have equal
dispersive coupling to two different resonant modes. In the first
version of our proposal described below, these two different modes
are realised as the fundamental modes of two different microwave resonators.
Crucial for the proposal is that each resonance should occur at nearly
(but not exactly) the same frequency. This near-coincidence allows
a scattering phase shift to wind through $2\times2\pi$ over a narrow
range of frequency. For the 4-qubit parity, a winding of $6\pi$ is
needed, requiring three closely spaced resonant modes to be involved.
Our second proposal will deal with this case. We will show that multiple
{\em resonances} need not require multiple {\em resonators}.
This proposal will use the currently popular ``3D\textquotedbl{}
cavity\cite{3D}, but one having a nearly cubical shape so that the
three lowest ${\mbox{{\rm TE}}}_{101}$-type modes\cite{RWvD} are
nearly degenerate. We will show a hypothetical scalable implementation
of the surface code within this scheme.





\section{Results -- 2D resonant structure}

We will first show that the necessary quantum-erasure function can
be realized by the choice of circuit hardware illustrated in Fig.
1. We will work through and discuss the case of the parity measurement
for three qubits; generalizing our construction to more qubits is
clear, and will be discussed below. In this construction we have two
resonators (we envision 1/4-wave coplanar waveguide resonators\cite{Pozar},
as illustrated schematically); in each we will employ one resonant
mode, with creation operators $a^{\dagger}$ and $b^{\dagger}$. It
would be normal to use the lowest-frequency (fundamental) modes; all
other modes are far separated in frequency (the next being at three
times the fundamental, in a 1/4-wave structure) and will be ignored.
In our scheme, these two fundamental frequencies $\omega_{a}$ and
$\omega_{b}$ should be almost, but not exactly, degenerate. The qubit-cavity
coupling will be the standard one given by the Jaynes-Cummings model
in the dispersive regime\cite{blais}. It will be non-standard only
in that each qubit will couple to both resonators, so that each qubit
$j$ will have a physical coupling strength $g_{j,a}$ to the $a$
resonator and $g_{j,b}$ to the $b$ resonator. The Hamiltonian of
the system of resonators and qubits can be written\cite{blaischi}
\begin{equation}
H=\left(\omega_{a}+\sum_{j=1}^{3}\chi_{a}^{j}\sigma_{z}^{j}\right)a^{\dagger}a+\left(\omega_{b}+\sum_{j=1}^{3}\chi_{b}^{j}\sigma_{z}^{j}\right)b^{\dagger}b+\frac{1}{2}\sum_{j=1}^{3}\omega_{j}\sigma_{z}^{j}.
\end{equation}
 Here $\omega_{j}$ are the qubit frequencies, and the dispersive
coupling parameters are $\chi_{a}^{j}=g_{j,a}^{2}/\Delta_{j,a}$ with
$\Delta_{j,a}=\omega_{j}-\omega_{a}$, and similarly for $\chi_{b}^{j}$
and $\Delta_{j,b}$. Note that we assume that the system should be
engineered so that there is no direct coupling between qubits.

\begin{figure}
\begin{centering}
\includegraphics[scale=0.3]{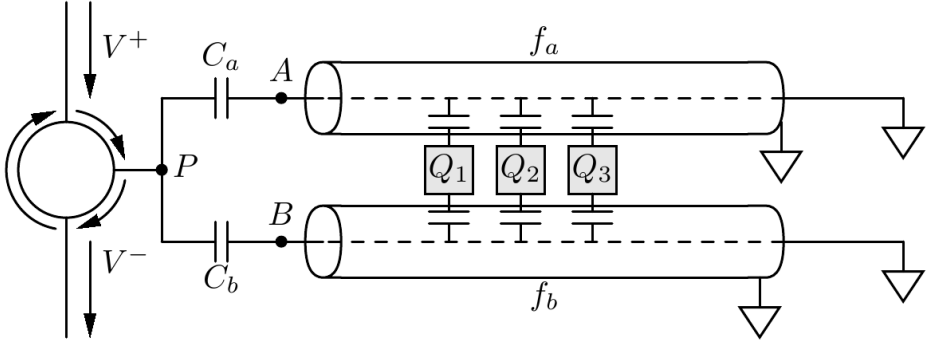} 
\par\end{centering}

\centering{}\caption{A schematic circuit QED setup for measuring the parity of three qubits.
Two transmission-line resonators have slightly different resonant
frequencies $f_{a}$ and $f_{b}$. The three qubits $Q_{1,2,3}$ (which
could be of the transmon type) each couple equally to the two resonators.
The parity information is contained in the phase of the reflection coefficient
at point $P$, which, throughout the action of the circulator, appears as a phase change of the
output signal $V^{-}$ relative to the imput tone $V^{-}$.}
\end{figure}

As indicated in Fig. 1, this resonator/qubit structure is to be coupled
capacitively to a scattering probe. Rather than extend our Hamiltonian
to include all these other details, we proceed in the following way:
Since $H$ commutes with each $\sigma_{z}^{j}$, we can examine the
Hamiltonian separately in each of its $2^{3}$ qubit eigensectors;
within each of these sectors $H$ describes a harmonic bosonic system,
with qubit-dependent resonant-frequency parameters. Thus, the full
scattering experiment can be described quite economically using the
classical language of impedance and scattering parameters; the conclusions
we draw from this classical discussion will have an immediate, standard
quantum interpretation in terms of coherent-state propagation.

From ordinary electrical transmission line theory, the impedance between
point $A$ in Fig. 1 and ground is given by 
\begin{equation}
Z_{A}(\omega)=iZ_{0}\tan\left(\frac{\pi}{2}\frac{\omega}{\omega_{r,a}}\right).
\end{equation}
 Here $Z_{0}=50\Omega$ is the impedance of the waveguide. The effective
resonant frequency $\omega_{r}$ is dependent on the state of the
three qubits $\left|s_{1}s_{2}s_{3}\right\rangle $ according to 
\begin{equation}
\omega_{r,a}=\omega_{a}+\sum_{j=1}^{3}(-1)^{s_{j}}\chi^{j}_a.
\end{equation}
 This same discussion applies to the qubit-state-dependent impedance
$Z_{B}(\omega)$ of the $B$ resonator. The impedance $Z_{P}(\omega)$
of the entire structure at point $P$ is then given by ordinary series-
and parallel-combination rules. In fact in the frequency range of
interest, the response is very well represented by the lumped
circuit of Fig. 2.

\begin{figure}[h]
\begin{centering}
\includegraphics[scale=0.3]{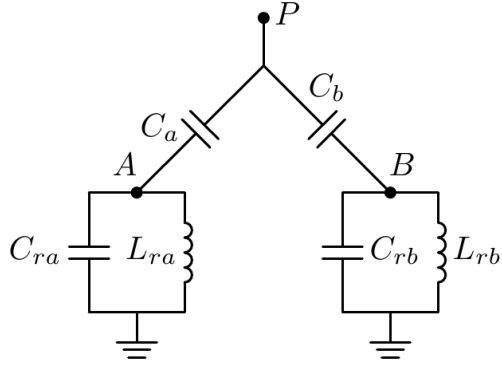} 
\par\end{centering}

\caption{Equivalent circuit for the double-resonator structure\cite{Pozar};
$C_{ra}=\frac{\pi}{4\omega_{r,a}Z_{0}}$ , $L_{ra}=\frac{1}{\omega_{r,a}^{2}C_{ra}}$
, $C_{rb}=\frac{\pi}{4\omega_{r,b}Z_{0}}$ , $L_{rb}=\frac{1}{\omega_{r,b}^{2}C_{rb}}$.}
\end{figure}

The measurable quantity for this structure is the reflection coefficient
$r$ at $P$. This is given by\cite{Pozar} 
\begin{equation}
r(\omega)=\frac{V^{-}(\omega)}{V^{+}(\omega)}=\frac{Z_{P}(\omega)-Z_{0}}{Z_{P}(\omega)+Z_{0}}.\label{eq:reflection-coefficient}
\end{equation}
 Note that because $Z_{P}$ is purely imaginary (lossless), $|r|=1$,
so that only the phase of $r$, 
\begin{equation}
\theta(\omega)\equiv\arg r(\omega)\label{eq:phase-response}
\end{equation}
 contains information (which can in fact be measured interferometrically).

Our object is to find a probe frequency $\omega_{p}$ such that the
reflected signals for all the even-parity qubit states, and all the
odd-parity states, are indistinguishable, but that the reflection
coefficient for the even and odd cases {\em are} distinct. This
will give us conditions on the reflected phase $\theta_{s_{1}s_{2}s_{3}}(\omega_{p})$
for the different qubit state settings. A general feature of the $\theta$
function will make this possible in our two-resonance setting. If
$\omega$ passes through a resonance of the system (pole of $Z_{P}$),
then, while the phase change of the impedance is $\pi$, the change
of the reflected phase is $2\pi$ (cf. Eqs. \eqref{eq:reflection-coefficient},
\eqref{eq:phase-response}). We will arrange that $Z_{P}$ has two
poles within a narrow range of frequency; this means that $\theta(\omega)$
will vary smoothly over $4\pi$ in that range. But, from the point
of view of a scattered tone, $\theta$ and $\theta+2\pi n$ are indistinguishable.
Thus, the $\theta(\omega)$ function varies over a sufficient range
that we can satisfy our quantum eraser condition for the parity measurement
in the following way: 
\begin{eqnarray}
\theta_{\mbox{\tiny even}} & \equiv & \theta_{000}(\omega_{p})=\theta_{011}(\omega_{p})+2\pi,\label{eq:even-parity-condition}\\
\theta_{\mbox{\tiny odd}} & \equiv & \theta_{111}(\omega_{p})=\theta_{001}(\omega_{p})-2\pi.\label{eq:odd-parity-condition}
\end{eqnarray}
 Since the $\chi$ coefficient for all qubits is taken to be equal,
the other necessary conditions, $\theta_{011}(\omega_{p})=\theta_{101}(\omega_{p})=\theta_{110}(\omega_{p})$
and $\theta_{001}(\omega_{p})=\theta_{010}(\omega_{p})=\theta_{100}(\omega_{p})$,
are satisfied automatically. It is also necessary that $\theta_{\mbox{\tiny even}}\neq\theta_{\mbox{\tiny odd}}$
(mod $2\pi$), with the best case (most distinguishable) being $\Delta\theta\equiv\theta_{\mbox{\tiny even}}-\theta_{\mbox{\tiny odd}}=\pi$.

One can show that for any choice of parameters in the two-pole circuit,
there exists a probe frequency $\omega_{p}$ and a dispersive shift
constant $\chi$ such that the parity-measurement conditions Eqs.
\prettyref{eq:even-parity-condition}, \prettyref{eq:odd-parity-condition}
are satisfied. However, if the resonant frequencies $\omega_{a}$
and $\omega_{b}$ are far apart compared with the width of the resonances,
$\theta_{\mbox{\tiny even}}-\theta_{\mbox{\tiny odd}}$ is very small.
By placing the resonances close to one another and choosing the capacitance
values carefully, a favorable solution can be found. We note that
for further optimization of this structure, there would be no difficulty
in replacing the simple pair of capacitors $C_{a,b}$ with a capacitance
bridge in a wye- or delta-configuration\cite{Pozar}.

\begin{figure}[t]
\begin{centering}
\includegraphics[scale=0.45]{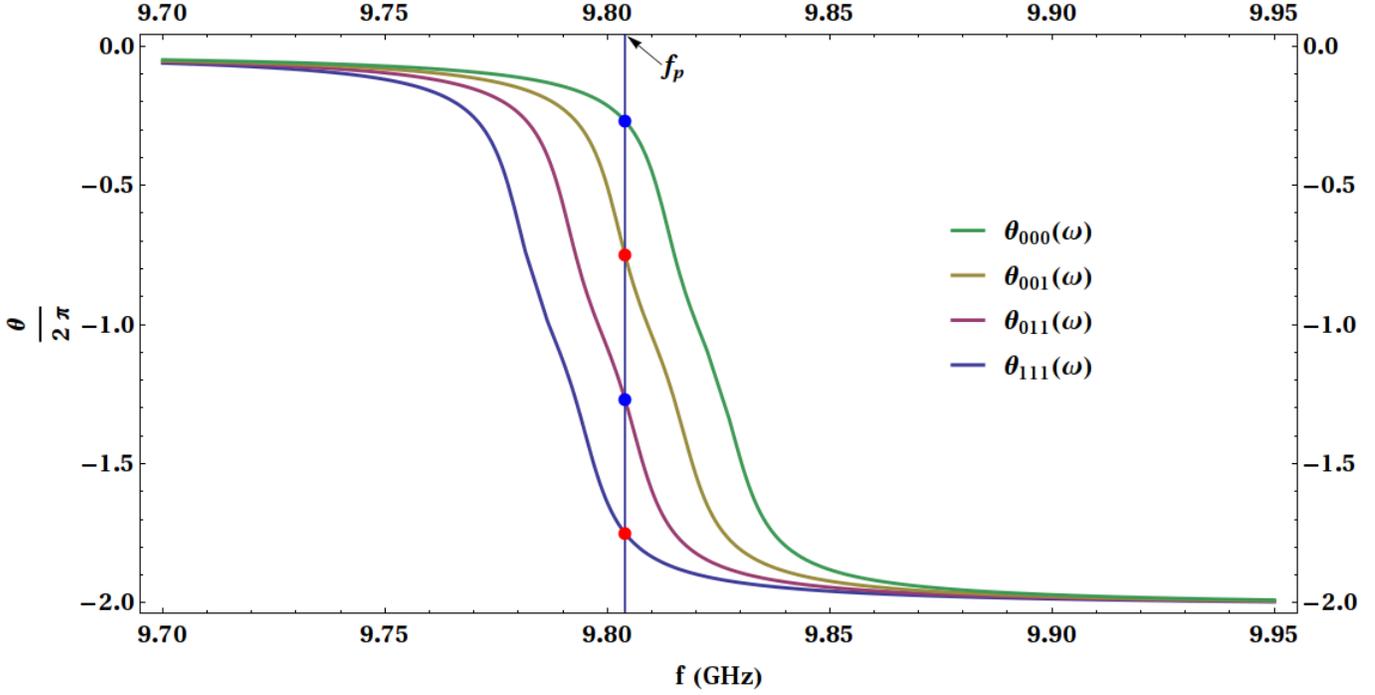} 
\par\end{centering}

\centering{}\caption{\label{fig:phase-dispersion-solution}Solution for realistic values
$\omega_{a}=2\pi(9.99{\mbox{ GHz}})$, $\omega_{b}=2\pi(10.01{\mbox{ GHz}})$,
$C_{a}=C_{b}=10\, fF$; with $\omega_{p}=2\pi(9.804\, GHz)$, $\chi=5.77\, MHz$,
giving $\Delta\theta=172.9^{\circ}$. Blue points correspond to even
states whereas red points correspond to odd states. Vertical blue
line shows the probe frequency $f_{p}=9.804\, GHz$.}
\end{figure}

Figure 3 shows a solution for the case of a realistic set of parameters.
The choices are $\omega_{a}=2\pi(9.99{\mbox{ GHz}})$, $\omega_{b}=2\pi(10.01{\mbox{ GHz}})$,
$C_{a}=C_{b}=10\, fF$; note that realistic values for coupling capacitors
are in $0.5-50fF$ range (see Ref. \cite{Wallraff}). Our Eqs. \prettyref{eq:even-parity-condition},
\prettyref{eq:odd-parity-condition} are satisfied (after a simple,
efficient numerical search) for $\omega_{p}=2\pi(9.804\, GHz)$ and
$\chi=5.77\, MHz$, giving the nearly optimal value $\Delta\theta=172.9^{\circ}$.
Note that the coupling capacitors will introduce $T_{1}$ qubit relaxation,
but we can estimate that capacitances on this scale give a cavity
loss rate of $\kappa\sim5{\mbox{MHz}}$. The Purcell formula for the
resulting qubit relaxation time is $T_{1(P)}=\frac{\Delta^{2}}{\kappa g^{2}}=\frac{\Delta}{\kappa\chi}$.
If we assume $\Delta=5{\mbox{GHz}}$, this gives $T_{1(P)}\sim200\mu s$.
Thus, the Purcell mechanism for relaxation will not be a severe limit
on the lifetime of the qubits.

We can qualitatively assess the result of applying the measurement
tone for a finite length of time. The signal-to-noise ratio for distinguishing
even from odd is largely a technical matter involving the noise performance
of amplifiers and the effective temperature of filters associated
with the resonator-qubit structure. The quantum-erasure property sets
a more fundamental limit. If the measurement time is $T$, the measurement
signal will then have a bandwidth $W\sim1/T$ around the probe frequency
$\omega_{p}$. Because the dispersion of the reflection response is
different for the different even and odd states (that is, $\frac{d\theta}{d\omega}$
is different for the distinct states, see Fig. \eqref{fig:phase-dispersion-solution}).
Thus we expect that to maintain the quantum-eraser condition, the
bandwidth $W$ should be kept to a small fraction of the resonance
width, so perhaps $T\sim10/\chi$. This gives a measurement time $T\sim2\mu s$.
While this is shorter than the expected $T_{1}$ in current devices,
it would be desirable to shorten $T$; we expect that further optimization
of the scattering structure could make all the $(\frac{d\theta}{d\omega})_{\mbox{\tiny odd}}$
and $(\frac{d\theta}{d\omega})_{\mbox{\tiny even}}$ more nearly equal,
so that perhaps $T$ could approach $1/\chi$.

A detailed calculation, given in the Appendix, confirms these qualitative
considerations. This calculation involves a quantum treatment of the
input tone $V^{+}$, in which it is written as a coherent state\cite{WM}
$\ket{\alpha}$, pulsed with a gaussian time profile with characteristic
time $T=1/W$. The pulse has mean photon number $|\alpha|^{2}$, and
therefore energy $\hbar\omega_{p}|\alpha|^{2}$. The output tone $V^{-}$
is also a coherent state $\beta_{{\bf s}}$, but dependent on the
qubit state $\ket{\bf s}=\ket{s_{1}s_{2}s_{3}}$. The coherent state
amplitude is always unchanged, $|\beta_{{\bf s}}|=|\alpha|$, but
it is dispersed differently for each state because of the scattering
phase shift.

The relevant result from the Appendix is 
\begin{equation}
\langle\beta_{{\bf s}}|\beta_{{\bf s'}}\rangle=1-\frac{|\alpha|^{2}b^{2}W^{2}}{2}+O(|\alpha|^{4}b^{4}W^{4}),\,\,\,\,\mbox{qubit parities {\bf s} and {\bf s'} the same},
\end{equation}
 
\begin{equation}
\langle\beta_{{\bf s}}|\beta_{{\bf s'}}\rangle=e^{-|\alpha|^{2}(1-\cos\Delta\theta)}\left(1+O\left(W\frac{d\theta}{d\omega}\right)\right),\,\,\,\,\mbox{qubit parities {\bf s} and {\bf s'} different}.
\end{equation}
 The new parameter $b$ is the first-order difference of phase dispersion,
\begin{equation}
b\equiv\theta_{{\bf s}}^{'}\left(\omega_{p}\right)-\theta_{{\bf s'}}^{'}\left(\omega_{p}\right).
\end{equation}
 The quantum-eraser condition requires that the ``same parity''
cases be indistinguishable ($\langle\beta_{{\bf s}}|\beta_{{\bf s'}}\rangle=1$),
and the ``different parity'' cases be perfectly distinguishable
($\langle\beta_{{\bf s}}|\beta_{{\bf s'}}\rangle=0$). Noting that,
on dimensional grounds, $b\sim1/\chi$, these quantum-erasure conditions
are well approximated so long as $T=1/W>|\alpha|b\sim|\alpha|/\chi$.
This confirms the qualitative discussion above, with the additional
insight that it is best if the probe is not too strong, i.e., if $\alpha$
is not too large (or course, it should be greater than one to satisfy
the ``different parity'' condition). If the photon number is taken
to be $|\alpha|^{2}=5$, and the pulse duration is $T\approx1\mu s$,
the peak pulse power will be about $P=\frac{\left|\alpha\right|^{2}\hbar\omega_{p}}{T}=-135\, dBm$.
This is indeed a weak signal, but in the detectable range with the
current state of the art\cite{Vijay-Siddiqi}.

\subsection{Comparison with Kerchhoff {\em et al.}}

\begin{figure}
\begin{centering}
\includegraphics[scale=0.25]{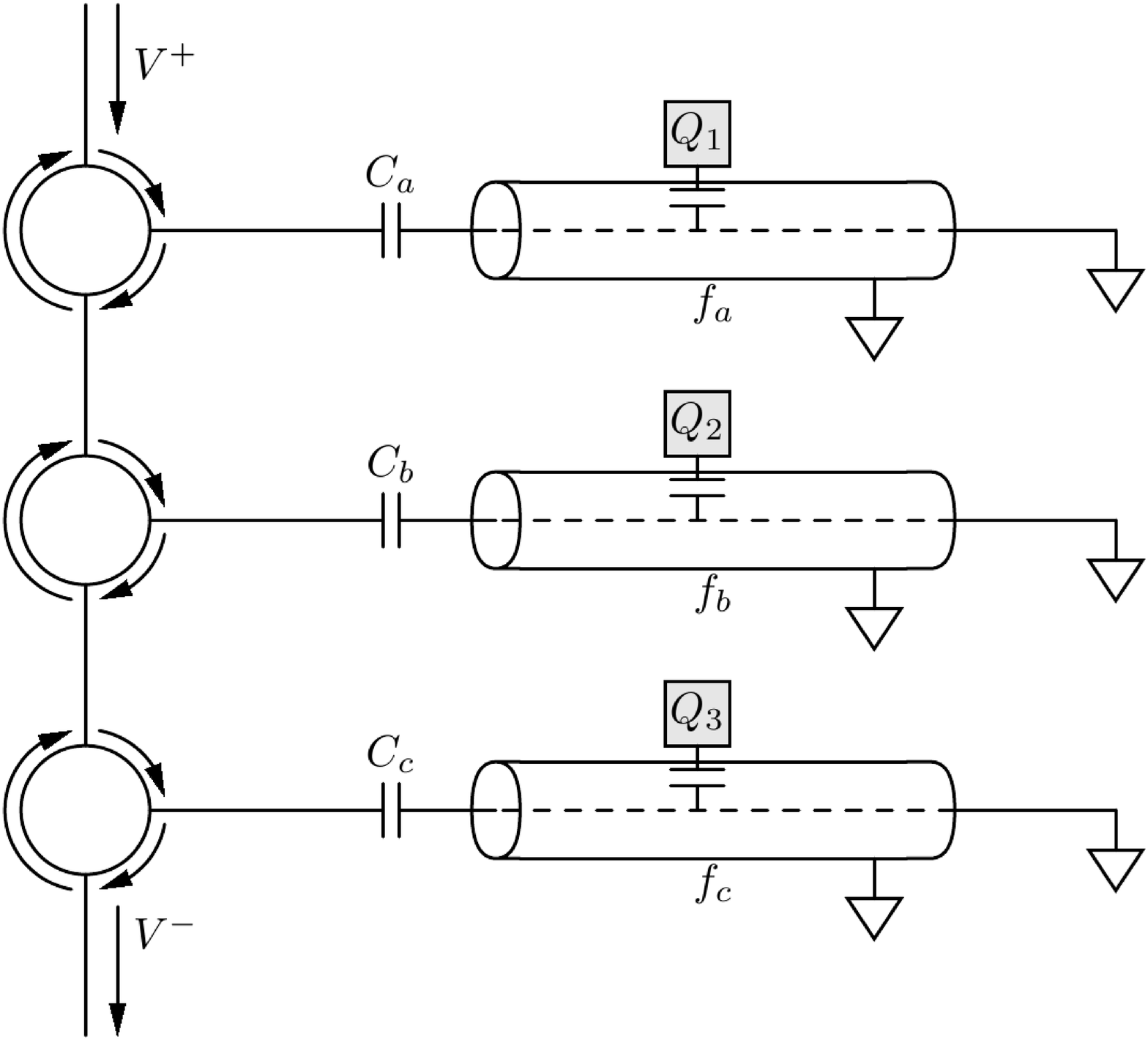} 
\par\end{centering}

\caption{Circuit for three-qubit extension of the scheme proposed as a cavity-optics
experiment in \cite{mabuchiway}, as it would be implemented with
microwave components in circuit QED.}
\end{figure}

Before presenting our second proposal, we can compare our concept
for three-qubit parity with the three-qubit extension of the two-qubit
parity measurement that was recently proposed by Kerchhoff, Bouten,
Silberfarb, and Mabuchi\cite{mabuchiway}. They propose sequential
scattering from a set of cavities, each coupled to one qubit; the
extension of their idea to three qubits, as rendered in plausible
microwave components, is shown in Fig. 4. Note that for $n$-qubit
parity this scheme uses $n$ resonators, while a generalisation of
our scheme above would use $\lceil(n+1)/2\rceil$ resonators (see
the following section). The Kerchhoff {\em et al.} scheme is certainly
elegant, and simpler in that each qubit only needs to couple to a
single cavity. The idea is to choose the coupling and the probe tone
such that there is a phase difference of $\pi$ between qubit state
0 and 1; then this network constructs a sum of the phases for each
qubit\cite{foot}. Each cavity should have exactly the same resonant
frequency (perhaps by tuning). This scheme is more obviously scalable
to more qubits than ours, and it would be possible in this scheme
to get rid of the first-order dispersion effects which degrade the
quantum-eraser condition, as calculated above. But, as we show in
the Appendix, the second-order dispersion difference, which would
still be present, leads to similar qualitative limits on the measurement
time and fidelity. Also, we point out that the quantum-eraser condition
would also be degraded by imperfections in the circulators, unlike
in our scheme. Given that there presently are no on-chip circulators
suitable for qubit experiments (but see progress in \cite{Bergeal,Koch}),
we believe that our scheme is closer to being realized with currently-available
components.

\section{Results -- 3D resonant structures}

The theoretical generalization of our scheme to the measurement of
the arguably more important case of 4-qubit parity is straightforward.
We require three closely spaced resonances $f_{a}$, $f_{b}$ and
$f_{c}$ , equal coupling $\chi$ of each qubit to each of the three
resonances, and a network that couples resonators to a single reflected
probe. In this case again all measurable quantities will emerge from
a single reflected-phase function $\theta_{{\bf s}}(\omega)$. Since
this function will vary smoothly over $6\pi$, it will be possible
to satisfy the three quantum-eraser equations for this case: 
\begin{eqnarray}
\theta_{\mbox{\tiny even}} & = & \theta_{0000}(\omega_{p})=\theta_{0011}(\omega_{p})+2\pi=\theta_{1111}(\omega_{p})+4\pi\\
\theta_{\mbox{\tiny odd}} & = & \theta_{0111}(\omega_{p})=\theta_{0001}(\omega_{p})-2\pi
\end{eqnarray}
 Tuning the $f_{i}$ values, $\chi$ and $\omega_{p}$ gives enough
freedom so that these equations should always be solvable. The analogous
conditions for $n$-qubit parity are straightforward to write down;
if we use the notation $\theta_{\mbox{w}t.i}$ to indicate the phase
shift if $i$ of the quibits are 1, then the condition is 
\begin{equation}
\theta_{\mbox{w}t.i}(\omega_{p})=\theta_{\mbox{w}t.i+2k}(\omega_{p})+2k\pi.
\end{equation}
 Combined with the desire that $\theta_{\mbox{w}t.0}(\omega_{p})\approx\theta_{\mbox{w}t.1}(\omega_{p})+\pi$,
we see that the $\theta$ function should vary by at least $\pi n$.
For even $n$ this will be accomplished with $n/2+1$ resonances (so
that $\theta$ winds through phase $\pi(n+2)$), while for odd $n$,
$(n+1)/2$ resonances suffice ($\theta$ winds through phase $\pi(n+1)$
in this case). For both even and odd $n$, the number of resonances
required can be written $\lceil(n+1)/2\rceil$, as stated above.

Returning to the consideration of 4-qubit parity, we see that some
new hardware elements would be required to achieve this by an extension
of the ``2D\textquotedbl{} (coplanar waveguide) scheme above. The
requirement that each qubit be equally coupled to each of three different
CPW resonators has not previously been achieved. It seems likely that
it is doable with the use of air bridges, which have only recently
entered the toolkit of quantum microwave engineering (see \cite{Simons,lehnert}).
Another possible way to obtain the multiple resonant modes needed
for multi-qubit parity measurement would involve the use of multi-conductor
CPWs\cite{Simons}, which naturally support modes that are closely
spaced in frequency.

But given these difficulties, we explore a second protocol to achieve
the implementation of 4-qubit parity measurement, involving ``3D''
superconducting cavities\cite{3D}. These high quality factor rectangular
cavities have recently been proven to be excellent implementations
of nearly decoherence-free Jaynes-Cummings physics\cite{newnigg}. In these the qubit-cavity
coupling is provided by antenna structures extending out from the
transmon qubits\cite{3D}. Multi-qubit structures\cite{bridge} have
been achieved in this technology, and bridging qubits, antenna-coupled
to two different cavities, are now possible\cite{bridgepriv,newnigg}. This
second protocol illustrates a further fact, which is that {\em multiple},
closely spaced resonances can be achieved within a {\em single}
resonant structure.

\begin{figure}
\begin{centering}
\includegraphics{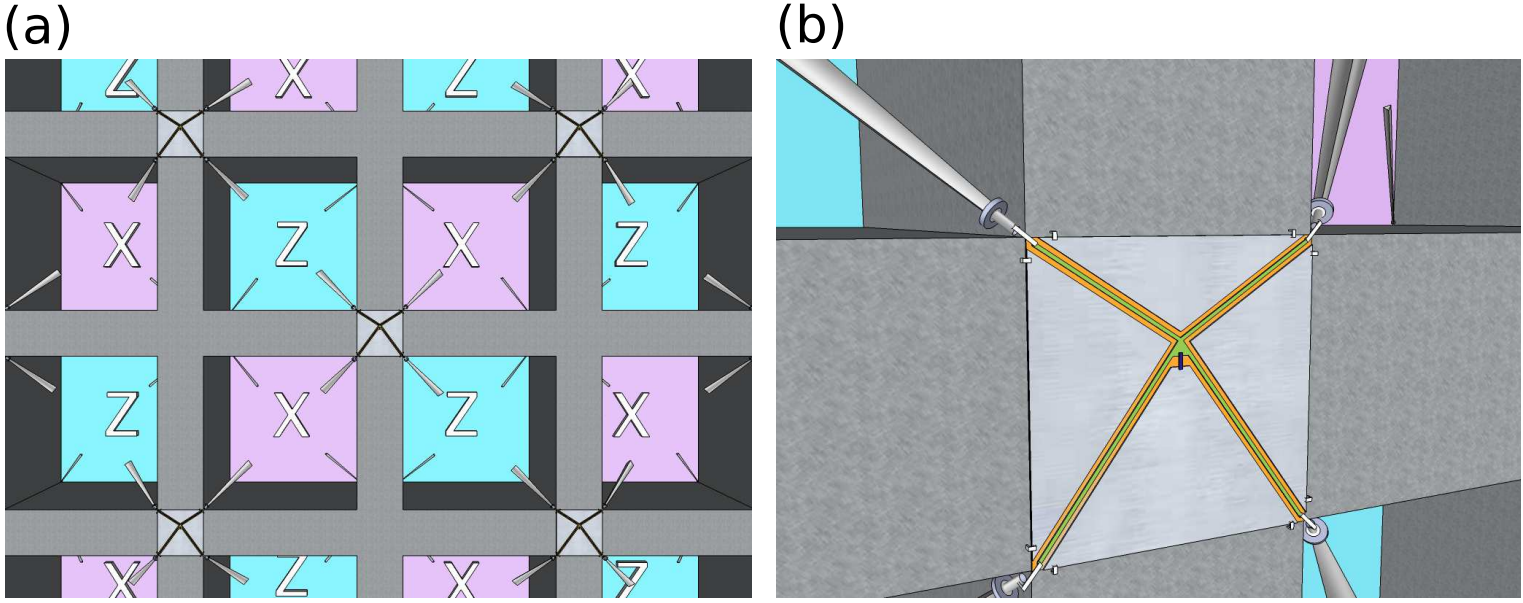}
\par\end{centering}

\raggedleft{}\caption{Potentially scalable implementation of the surface code using direct
4-qubit parity measurements\cite{sket}. a) Overview of the proposed structure.
A solid plate of metal (Al or Cu, as in recent experiments\cite{3D,RigettiCu})
is perforated with nearly square holes. The thickness of the plate
is close to the dimension of the square. When closed with thin top
and bottom plates (not shown), these holes become nearly cubical microwave
cavities. Qubit structures are grown on chips (lighter squares) using
existing thin-film techniques. The insulating side of the chip is
mounted to a junction point in the perforated plate; half of the junctions
contain chips at the front surface of the plate, and the other half
are on the back side (not visible). Each qubit is in electrical contact,
via leads going to the corners of the chip, with four antenna structures
(tapered rods) projecting diagonally into the four cubical cavities
surrounding the chip. Each cavity will enable a four qubit parity
measurement, either ZZZZ or XXXX, as indicated. The external feed
lines needed to interrogate each cavity by reflectometry will enter
via one of the unoccupied corners of the cubes (not shown). b) Close-up
around one qubit chip. It could be made in the conventional way, in
which a thin film of metal is etched away in slot regions, exposing
the underlying insulating substrate (orange). This forms four short
segments of coplanar waveguide structures, with the center conductor
(green) going to each of the corners of the chip. One small structure
(blue) containing a Josephson junction connects the central region
of center conductor to one of the ground planes (light grey). The
four triangular ground-plane regions are tied to the metal plate,
and therefore to each other, by bridges or wire bonds near the corners
of the chips. Such bridge/bond connections are also made between the
four center conductors and the antenna rods. Insulating support structures
isolating the antennas from the conducting walls are only schematically
shown. The consequence of this connection arrangement is that the
Josephson junction is shunted by four capacitances in series, each
formed by an antenna and the side walls of the cavity containing it. Presumed
optimal dimensions for the chip (sub-millimeter) and cavity (few centimeter)
are not accurately depicted.}
\end{figure}

Figure 5 illustrates the concept of this second protocol, also showing
how it could be extended to be part of a scalable implementation of
the surface code architecture\cite{DDV,whatssurface}. The key modification
in this structure is that the 3D cavities, rather than having a large
aspect ratio in which $w>>l>>h$ (w=width, l=length, h=height), should
be nearly, but not exactly, cubical. Thus, the three lowest resonant
modes would be nearly degenerate in frequency, since their wavelengths
will be set by $\pi/w$, $\pi/l$, and $\pi/h$. (Be warned that the
traditional labels for these modes are ${\mbox{{\rm TE}}}_{101}$,
${\mbox{{\rm TE}}}_{011}$, and ${\mbox{{\rm TM}}}_{110}$ \cite{RWvD}.)
Each cavity is coupled, via antenna structures, to four qubits, two
on the front surface in Fig. 5 and two on the back (not shown). Each
antenna runs along the body diagonal (i.e., $\langle111\rangle$ axis)
of the cubical cavity; with this geometry the coupling is equal to
each of the three eigenmodes, whose modal electric field patterns
point straight along one of the three coordinate axes.

Unlike for our first (``2D'') proposal, we will not provide calculations
of the performance of the structure of Fig. 5. While the structure
of Fig. 1 can be analyzed using elementary transmission line theory
and a few component parameter values whose ranges are well understood,
the system of Fig. 5 is a complex, three-dimensional structure
whose electromagnetic response can only be obtained reliably by detailed
calculations that are beyond the scope of this paper. It is encouraging
that progress is being made in the detailed modeling of couplings
in such a 3D geometry\cite{blackbox}. We note that many details of
Fig. 5, such as the thin-film metallisation around the qubit, contacts
between chip conductors and antenna and cavity metal, the shape of
the antennas, and the exact geometry of the cavities, should be optimised
by detailed simulation. One comment on the cavity structure: while
the cubical cavity satisfies the requirement of three closely spaced
modes in an elegant way, it will be perhaps discouragingly large (centimeter-scale)
from the point of view of potential scale-up of the surface code.
Work is commencing on much different forms of resonator geometries,
which by being ``quasi-lumped''\cite{quasilumped}, can be much
more physically compact. It will not be necessary to remain only with
the Platonic cavities.

\section{Discussion and Conclusions}

We consider our basic schemes of Figs. 1 and 5 to be realistic for
implementation by experiment. For Fig. 1, a precise thin-film layout
of an on-chip structure (to the right of point $P$ in Fig. 1) could
be devised based on the parameter values we have determined. It is
clear that detailed electromagnetic modelling would be useful to guide
the layout design\cite{threeport}. The values obtained for $C_{a}$
and $C_{b}$ correspond to well-known few-finger interdigitated capacitors.
We foresee three main difficulties: 1) Coupling each qubit to the
two resonators. 2) Tuning the qubits to achieve the equal-$\chi$
condition. The topology of the chip layout would preclude bringing
flux-bias lines to each qubit; it is possible that only two of the
three qubits would need to be tuned, and one of these tunings could
be from an external magnetic field. 3) Measuring few-photon signal
levels. It appears that the standard HEMT amplifier arrangements would
have too much noise for the conditions we have calculated; adoption
of new, quantum-limited amplification will be needed for high-fidelity
readout.

The 3D scheme of Fig. 5 will require more work to assess its optimal
implementation. Achieving the parameter values envisioned by our theory
will require detailed simulation of the complex structure that we
have proposed, with many details subject to variation. Since the direct,
ancillaless approach studies here is so different from the existing
technique of collecting error syndromes with the action of a quantum
circuit using ancillas, we believe that it is impossible to know which
approach will be superior. Further studies and refinement of both
approaches will be necessary to determine which will provide the best
way forward.

So, it may be that our proposals can only attain satisfactory performance
in structures in which the resonators are tuneable (as in \cite{delsing,CEAtune,LehnertTune}),
or qubits are tuneable in frequency (as in \cite{2bitalg} and in
many other works) or in effective coupling strength (as in \cite{tunq}).
Application of these device construction techniques, as well as near-quantum
limited amplification\cite{Vijay-Siddiqi}, will be likely be needed
to achieve high-fidelity, single-shot parity measurement as envisioned
in he proposals we give here. We hope that following this route will
indeed be facilitated by the many interesting experimental\cite{goodcomponents,1qubit2res,Hoi}
and theoretical\cite{morecomponents,mphoton,jmeastheory} innovations
in the application of circuit QED that we see presently. 

The proposals of this paper are not the blueprint for a scalable quantum
computer; they are concepts on which further detailed studies to determine
optimal device functionality can be based. While the structures that
we suggest here will by no means be trivially realised, we believe
that the crucial role played by parity measurement in the implementation
of reliable quantum computation makes these approaches worthwhile
to pursue.

Thanks to Barbara
Terhal for emphasising the importance of multi-qubit parity measurements
for fault-tolerant quantum computation; we thank Jay Gambetta, John
Smolin, and especially Bob W. Newcomb, for ongoing discussions. We
are grateful for support from the Alexander von Humboldt foundation.

\section{Appendix}

In this Appendix we will compute fidelities between output signals
for the case of a finite bandwidth probe.

We compute the fidelity between output signals corresponding to two
different qubit states $\ket{\bf s}$ and $\ket{\bf s'}$. We first
construct the creation operator $\hat{a}_{pulse}^{\dagger}$ corresponding
to a finite bandwidth probe pulse as a linear combination of a densely
spaced, discrete set of harmonic modes $\hat{a}_{\omega_{i}}^{\dagger}$
\cite{glauber}:

\begin{equation}
\hat{a}_{pulse}^{\dagger}=\underset{i}{\sum}C_{i}\hat{a}_{\omega_{i}}^{\dagger}
\end{equation}
 with

\begin{equation}
C_{i}=\frac{\sqrt{\delta\omega}e^{-\left(\omega_{i}-\omega_{p}\right)^{2}/4W^{2}}}{\left(2\pi W^{2}\right)^{1/4}},
\end{equation}
 where $\omega_{p}$ is the center frequency, $W$ the bandwidth of
the probe signal, $\omega_{i}$ is the frequency of mode $i$, and
$\delta\omega$ is the difference between frequencies of successive
modes. In the above expression the weight $C_{i}$ of each term is
chosen such that in the limit of a continuum of modes ($\delta\omega\rightarrow0$)
the unitarity constraint is satisfied\cite{glauber}:

\begin{eqnarray}
{\sum}\left|C_{i}\right|^{2} & \approx & \frac{1}{\sqrt{2\pi W^{2}}}\int_{-\infty}^{\infty}d\omega e^{-\left(\omega-\omega_{p}\right)^{2}/2W^{2}}\\
 & = & 1+O(\delta\omega).
\end{eqnarray}
 Now we define a coherent state of amplitude $\alpha$ for probe pulse:

\begin{equation}
\left|\alpha\right\rangle =e^{-\left|\alpha\right|^{2}/2}e^{\alpha\hat{a}_{pulse}^{\dagger}}\left|0\right\rangle .\label{eq:coherent-pulse-state}
\end{equation}
 If we define an amplitude $\alpha_{i}$ for mode $i$ as follows:

\begin{equation}
\alpha_{i}\equiv\alpha C_{i}=\alpha\frac{\sqrt{\delta\omega}e^{-\left(\omega_{i}-\omega_{p}\right)^{2}/4W^{2}}}{\left(2\pi W^{2}\right)^{1/4}}\label{eq:alphai}
\end{equation}
 using the fact

\begin{equation}
\underset{\delta\omega\rightarrow0}{\lim}\underset{i}{\sum}\left|\alpha_{i}\right|^{2}=\left|\alpha\right|^{2}\underset{\delta\omega\rightarrow0}{\lim}\underset{i}{\sum}\left|C_{i}\right|^{2}=\left|\alpha\right|^{2}
\end{equation}
 we can rewrite the coherent probe state $\left|\alpha\right\rangle $
in Eq. \eqref{eq:coherent-pulse-state} in the continuum modes limit
as:

\begin{align}
\left|\alpha\right\rangle  & =e^{-\left|\alpha\right|^{2}/2}e^{\alpha\hat{a}_{pulse}^{\dagger}}\left|0\right\rangle \\
 & =e^{-\frac{1}{2}\underset{i}{\sum}\left|\alpha_{i}\right|^{2}}e^{\alpha\underset{i}{\sum}C_{i}\hat{a}_{\omega_{i}}^{\dagger}}\left|0\right\rangle \\
 & =\underset{i}{\prod}e^{-\left|\alpha_{i}\right|^{2}/2}e^{\alpha_{i}\hat{a}_{\omega_{i}}^{\dagger}}\left|0\right\rangle \\
 & =\underset{i}{\prod}\left|\alpha_{i}\right\rangle 
\end{align}
 where we have defined a coherent state for each mode $i$ as

\begin{equation}
\left|\alpha_{i}\right\rangle =e^{-\left|\alpha_{i}\right|^{2}/2}e^{\alpha_{i}\hat{a}_{\omega_{i}}^{\dagger}}\left|0\right\rangle .
\end{equation}
 Now if the qubits are in state $\ket{\bf s}$ the coherent component
$\left|\alpha_{i}\right\rangle $ of the input probe at frequency
$\omega_{i}$ will get a phase shift of $\theta_{{\bf s}}\left(\omega_{i}\right)$
and go to the state $\left|\alpha_{i}e^{i\theta_{{\bf s}}\left(\omega_{i}\right)}\right\rangle $.
If we call $\left|\beta_{{\bf s}}\right\rangle $ the state of the
output signal when qubits are in state $\left|{\bf s}\right\rangle $
we can compute the fidelity

\begin{align}
\mathcal{F} & =\left\langle \beta_{{\bf s}}|\beta_{{\bf s'}}\right\rangle \\
 & =\underset{i}{\prod}\left\langle \alpha_{i}e^{i\theta_{{\bf s}}\left(\omega_{i}\right)}|\alpha_{i}e^{i\theta_{{\bf s'}}\left(\omega_{i}\right)}\right\rangle \\
 & =\underset{i}{\prod}\exp\left\{ -\left|\alpha_{i}\right|^{2}(1-e^{-i\left(\theta_{{\bf s}}\left(\omega_{i}\right)-\theta_{{\bf s'}}\left(\omega_{i}\right)\right)})\right\} \\
 & =\exp\left\{ -\underset{i}{\sum}\left|\alpha_{i}\right|^{2}(1-e^{-i\left(\theta_{{\bf s}}\left(\omega_{i}\right)-\theta_{{\bf s'}}\left(\omega_{i}\right)\right)})\right\} .
\end{align}
 Expanding the phases around the center frequency of the probe $\omega_{i}=\omega_{p}+\Delta\omega_{i}$

\begin{align}
\theta_{{\bf s}}\left(\omega_{i}\right) & =\theta_{{\bf s}}\left(\omega_{p}\right)+\theta_{{\bf s}}^{'}\left(\omega_{p}\right)\Delta\omega_{i}+O\left(\left(\Delta\omega_{i}\right)^{2}\right),\\
\theta_{{\bf s'}}\left(\omega_{i}\right) & =\theta_{{\bf s'}}\left(\omega_{p}\right)+\theta_{{\bf s'}}^{'}\left(\omega_{p}\right)\Delta\omega_{i}+O\left(\left(\Delta\omega_{i}\right)^{2}\right).
\end{align}
 Now, we consider the case when \textbf{s} and \textbf{s'} have the
same parity. Then $\theta_{{\bf s}}\left(\omega_{p}\right)=\theta_{{\bf s'}}\left(\omega_{p}\right)\mbox{ mod(2\ensuremath{\pi})}$,
and we get to the first order in $\Delta\omega_{i}$

\begin{equation}
\mathcal{F}=\exp\left\{ -\underset{i}{\sum}\left|\alpha_{i}\right|^{2}(1-e^{-ib\Delta\omega_{i}})\right\} 
\end{equation}
 where we have defined 
\begin{equation}
b\equiv\theta_{{\bf s}}^{'}\left(\omega_{p}\right)-\theta_{{\bf s'}}^{'}\left(\omega_{p}\right).
\end{equation}
 Using the expression for $\alpha_{i}$ in Eq. \eqref{eq:alphai}
and taking the limit of continuum of modes the sum in the exponent
of the above expression becomes an integral

\begin{align}
\mathcal{F} & =\exp\left\{ -\frac{\left|\alpha\right|^{2}}{\sqrt{2\pi W^{2}}}\int_{-\infty}^{\infty}d\omega e^{-\left(\omega-\omega_{p}\right)^{2}/2W^{2}}(1-e^{-ib\left(\omega-\omega_{p}\right)})\right\} \\
 & =\exp\left\{ -\left|\alpha\right|^{2}\left(1-e^{-b^{2}W^{2}/2}\right)\right\} .
\end{align}
 If $bW\ll1$ then $1-e^{-b^{2}W^{2}/2}\simeq\frac{b^{2}W^{2}}{2}$
so that

\begin{equation}
\mathcal{F}\simeq e^{-\left|\alpha\right|^{2}b^{2}W^{2}/2}.
\end{equation}
 If we further assume that $\left|\alpha\right|bW\ll1$ we get

\begin{equation}
\mathcal{F}\simeq1-\frac{\left|\alpha\right|^{2}b^{2}W^{2}}{2}.
\end{equation}

The fidelity between odd and even states is given by a simpler calculation:
\begin{align}
\mathcal{F}_{even/odd} & =\left\langle \alpha e^{i\theta_{\mbox{\tiny even}}}|\alpha e^{i\theta_{\mbox{\tiny odd}}}\right\rangle \\
 & =e^{-\left|\alpha\right|^{2}(1-\cos\Delta\theta)}\approx e^{-2\left|\alpha\right|^{2}}.
\end{align}
 The final expression is a consequence of the fact that $\Delta\theta\approx\pi$.

Another case we look at, relevant for the alternative scheme of \cite{mabuchiway},
is the case of matching linear dispersion (i.e. $b=0$) but finite
quadratic dispersion mismatch

\begin{equation}
b'\equiv\left.\frac{d^{2}\theta_{{\bf s}}\left(\omega\right)}{d\omega^{2}}\right|_{\omega_{p}}-\left.\frac{d^{2}\theta_{{\bf s}}\left(\omega\right)}{d\omega^{2}}\right|_{\omega_{p}}.
\end{equation}
 Then the fidelity, for the case of the same parity, is

\begin{equation}
\mathcal{F}=\left|\exp\left\{ -\underset{i}{\sum}\left|\alpha_{i}\right|^{2}(1-e^{-ib'\left(\Delta\omega_{i}\right)^{2}})\right\} \right|.
\end{equation}
 Again in the limit of continuum of modes $\mathcal{F}$ becomes 
\begin{align}
\mathcal{F} & =\left|\exp\left\{ -\frac{\left|\alpha\right|^{2}}{\sqrt{2\pi W^{2}}}\int_{-\infty}^{\infty}d\omega e^{-\left(\omega-\omega_{p}\right)^{2}/2W^{2}}(1-e^{-ib'\left(\omega_{i}-\omega_{p}\right)^{2}})\right\} \right|\\
 & =\left|\exp\left\{ -\left|\alpha\right|^{2}\left(1-\frac{1}{\sqrt{1+2ib'W^{2}}}\right)\right\} \right|\\
 & =\exp\left\{ Re\left[-\left|\alpha\right|^{2}\left(1-\frac{1}{\sqrt{1+2ib'W^{2}}}\right)\right]\right\} \\
 & =\exp\left\{ -\left|\alpha\right|^{2}\left(1-\sqrt{\frac{1+\sqrt{1+4b'^{2}W^{4}}}{2+8b'^{2}W^{4}}}\right)\right\} \\
 & \simeq1-\frac{3\left|\alpha\right|^{2}b'^{2}W^{4}}{2}+O\left(\left|\alpha\right|^{4}\left(b'W^{2}\right)^{4}\right)
\end{align}
 where we assumed that $b'W^{2}\ll1$ and $\left|\alpha\right|b'W^{2}\ll1$.

To get an estimate for the power of the measurement signal we assume
that the signal contains $\left|\alpha\right|^{2}=5$ photons and
that it has duration $T=1\mu s$ so that the peak power $P$ will
roughly be $P=\frac{\left|\alpha\right|^{2}\hbar\omega_{p}}{T}=-135\, dBm$.

\end{document}